\documentclass[preprint,12pt]{elsarticle}

\usepackage{amssymb}
\usepackage{amsmath}

\usepackage[table, svgnames]{xcolor}
\usepackage{graphicx}
\usepackage{dcolumn}
\usepackage{bm}
\usepackage{multirow} 
\usepackage{color}
\usepackage{graphicx} 
\usepackage{mathtools}
\usepackage{multirow}
\usepackage{caption}
\usepackage{amsmath} 
\usepackage{chemformula} 
\usepackage{csquotes} 
\biboptions{numbers,sort&compress}
\usepackage{subfig}
\usepackage{placeins}
\usepackage{cancel}
\usepackage[normalem]{ulem}
\usepackage{siunitx}
\usepackage{bm}
\usepackage[capitalise]{cleveref}
\usepackage{booktabs}
\usepackage{xr}
\usepackage{comment}
\usepackage{csvsimple}
\usepackage{breqn}

\usepackage{amsmath}

\journal{Advanced Engineering Informatics}

\begin{document}

\begin{frontmatter}



\title{Latent Space Diffusion for Topology Optimization}


\author[A]{Aaron Lutheran}
\author[B]{Srijan Das}
\author[A,C]{Alireza Tabarraei\corref{mycorrespondingauthor}}
\cortext[mycorrespondingauthor]{Corresponding author}
\ead{atabarra@charlotte.edu} 

\address[A]{Department of Mechanical Engineering and Engineering Science, The University of North Carolina at Charlotte, Charlotte, NC 28223, USA}
\address[B]{Department of Computer Science, The University of North Carolina at Charlotte, Charlotte, NC 28223, USA}
\address[C]{School of Data Science, The University of North Carolina at Charlotte, Charlotte, NC 28223, USA}


\begin{abstract}
Topology optimization enables the automated design of efficient structures by optimally distributing material within a defined domain. However, traditional gradient-based methods often scale poorly with increasing resolution and dimensionality due to the need for repeated finite element analyses and sensitivity evaluations. In this work, we propose a novel framework that combines latent diffusion models (LDMs) with variational autoencoders (VAEs) to enable fast, conditional generation of optimized topologies. Unlike prior approaches, our method conditions the generative process on physically meaningful fields—specifically von Mises stress, strain energy density, volume fraction, and loading information—embedded as dense input channels. To further guide the generation process, we introduce auxiliary loss functions that penalize floating material, load imbalance, and volume fraction deviation, thereby encouraging physically realistic and manufacturable designs. Numerical experiments on a large synthetic dataset demonstrate that our VAE-LDM framework outperforms existing diffusion-based methods in compliance accuracy, volume control, and structural connectivity, providing a robust and scalable alternative to conventional topology optimization techniques for complex design tasks.
\end{abstract}


\end{frontmatter}



\section{\textbf{Introduction}}\label{introduction}		
Advancements in additive manufacturing techniques have revolutionized the production and application of complex structural designs, diverging significantly from traditional design methodologies. Additive manufacturing techniques allow for the creation of intricate features, such as organic shapes, hollow components, and lattice structures, which enable the creation of more efficient and effective designs \cite{Thompson2016, Rosen2017, kuo2018support}. Topology optimization is a design optimization approach that allows for the complex feature set that additive manufacturing is capable of producing. The topology optimization scheme discretizes the physical domain into small finite elements. These elements each have an assigned density that determines their response to load. Element densities are the parameters of the topology optimization design, and are updated with minimum finding algorithms. Structuring the problem in this way enables the algorithm to generate complex freeform structures when discretized with a fine enough mesh resolution \cite{rozvany_generalized_1992}.

Traditional methods for topology optimization, such as density-based \cite{bendsoe_optimal_1989} and level-set approaches \cite{allaire_level-set_2002}, often struggle with computational efficiency. Topology optimization algorithms are usually gradient based approaches that require iteratively calculating the compliance objective repeatedly until it converges to an approximate solution. The process also requires computationally expensive sensitivity analysis, which must be performed at each step \cite{andreassen_efficient_2011}. Because the update in topology optimization is performed on every pixel in a design domain, the computational complexity dramatically increases as the resolution of the domain is increased. When expanded into 3D, even modest mesh resolutions can lead to domains that take hours to optimize on. This compromise, of runtime with resolution, has prompted exploration in the use of machine learning techniques to improve the optimization algorithms.

Machine learning methods offer opportunities to increase computational efficiency of topology optimization methods by avoiding the need tor the complex sensitivity calculations that traditional topology optimization requires. Deep neural networks have demonstrated initial success in structural optimization as well as coupled thermo-mechanical topology optimization problems \cite{shishir_multimaterials_2024, tabarraei2025variational}. These techniques use the position of an element as the input, requiring a sampling process over all elements to generate an output. This method produces good results, but the generation of a single sample requires an iterative training process at runtime. Recent advancements in machine learning, particularly in image generation models like generative adversarial networks (GANs) \cite{nie_topologygan_2021}, and denoising diffusion probabilistic models (DDPMs), have offered new opportunities to address these challenges \cite{oh_deep_2019, maze_diffusion_2022}. These techniques generate the topology as an image, requiring a single forward pass or, for diffusion models, a preset chain of forward passes. While effective, these models have their own drawbacks. GANs produce results quickly and effectively, but require a delicately balanced training process between the generator and discriminator models, leading to a higher likelihood of collapse before training is complete. Diffusion models have been demonstrated to achieve higher performance than GANs \cite{maze_diffusion_2022},  but require a complex sampling process in an inefficient image diffusion process.

This paper expands on the work of TopoDiff, \cite{maze_diffusion_2022} a conditional diffusion model approach to topology optimization, by replacing the architecture with a latent diffusion model approach. Incorporating an autoencoder model in the diffusion architecture improves the efficiency of the de-noising model and simplifying the de-noising process. The autoencoder serves to reduce the dimensionality of the image space, reducing the number of parameters and computations required for sampling. This approach also gives an opportunity to engineer the latent space to accommodate our particular use case, with the use of a variational auto encoder.

Variational autoencoders (VAEs) perform the same semantic compression that an autoencoder does, with the stipulation that the latent space is sampled from a normal distribution. This sampling step ensures that the latent space learned by the autoencoder performs a blending between nearby topologies. The sampling step ensures that the latent space has continuity between two encoded topologies. VAEs are also formulated with a loss component to the magnitude of the latent vector, which encourages representations to be nearby in the latent space. The diffusion model is trained to learn a distribution from a noisy sample to the distribution of expected topologies. Using a VAE for the latent space will provide the diffusion model with a more complete and continuous distribution to map to.

The reference architecture, TopoDiff, utilizes a pair of informing models that are trained on predicting the presence of floating material and on predicting the approximate compliance of the structure. This information is used to inform the posterior prediction of the diffusion model, which requires the models to be robust to noise. We propose an auxiliary loss metric that directly incorporates the floating material property, as well as determining load discrepancy and volume fraction error. Avoiding the training of these auxiliary models speeds up the training process, while also removing a potential source of error from informing models that need to predict floating material and compliance from a noisy image. Applying this loss metric to the new VAE model will condition the latent representation to prefer representations that abide by these constraints. We find that the compliance objective is necessarily minimized in the de-noising process, mitigating the need for a compliance predicting auxiliary model.

\section{\textbf{Topology Optimization}}\label{sec:02}

\begin{figure}
    \centering
    \includegraphics[width=0.85\linewidth]{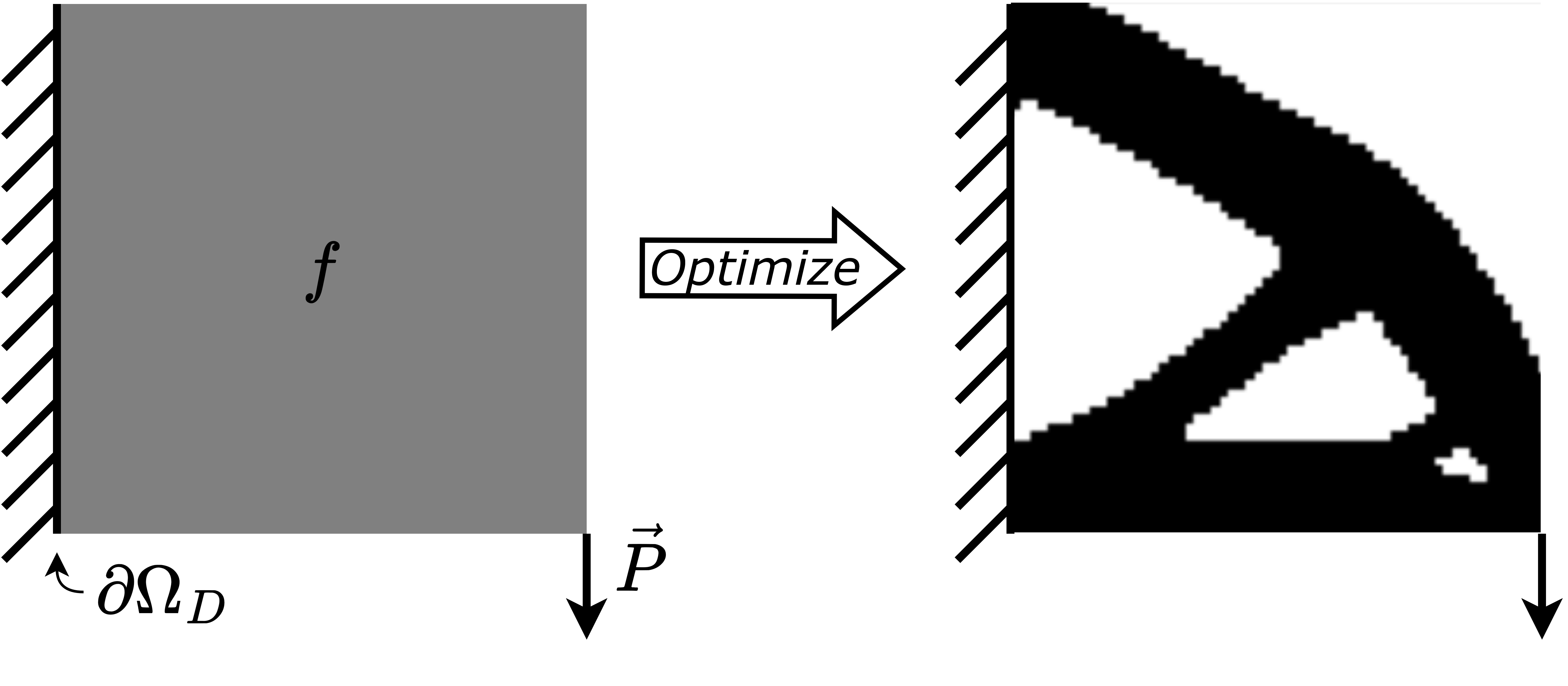}
    \caption{Example of the topology optimization process, for a specific combination of boundary condition $\partial\Omega_D$ and load $P$}
    \label{fig:Topopt}
\end{figure}

The aim of Topology Optimization is to find the optimal material distribution within a design domain that best satisfies a given objective function while adhering to constraints such as weight, stress, or displacement limits. \cref{fig:Topopt} demonstrates the typical optimization problem formulation and respective optimized solution. An open domain with some volume fraction $f$ is to be designed on, while subjected to a force vector $\overrightarrow{P}$ and fixed boundary $\partial\Omega_D$. For our study, we use the solid isotropic material penalization (SIMP) formulation of structural topology optimization for generating our dataset. SIMP is widely used due to its simplicity and effectiveness in enforcing a material distribution of solid or void regions within the design space.

The SIMP method works by discretizing the design domain into elements for Finite Element Analysis (FEA). Each element is assigned a density value, $x_e$, which relates to that element's Young's Modulus according to the penalization function:

\begin{equation} 
    E_e(x_e) = E_{min} + x_e^p (E_0-E_{min})
    \label{Penalization} 
\end{equation}

where $E_0$ is the elastic modulus of the material and $E_{min}$ is a small stiffness $(~0.001E_0)$ to prevent errors. The penalization factor, $p$, is introduced to encourage a binary material distribution (either solid or void) by discouraging intermediate density values. The design variable $x_e$ is clamped between $0$ and $1$, ensuring that the possible Young's Modulus values $E_e(x_e)$ are within the range [$E_{min}$, $E_0$].

For structural topology optimization, the compliance of the structure, defined as the total strain energy in the system, is taken as the objective function Eq. \eqref{Compliance}. Due to the penalization function Eq. \eqref{Penalization}, lower compliance structures correspond to structures with more material due to the lower deformation that occurs under loading. To prevent fully solid structures, a desired volume fraction constraint $f$ is applied. The optimization problem is then defined as:
\begin{eqnarray}
\min_{x} c(x) = \mathbf{U}^T \mathbf{KU} \label{Compliance}\\
\mathbf{KU} = \mathbf{F} \label{FEA}\\
V(x)/V_0 \leq f \label{Vol Frac}\\
0 \leq x_e \leq 1 \label{Clamp}
\end{eqnarray}
where $\mathbf{F}$ is the global force vector, $\mathbf{U}$ is the displacement vector obtained from FEA. $V(x)$ is the volume of the topology, $V_0$ is the total volume of the domain, and $f$ is the prescribed maximum volume fraction.

This formulation ensures that material is distributed efficiently within the design space to achieve optimal stiffness while adhering to the given volume constraints. The optimization process iteratively updates the element densities until convergence is reached, leading to a design that maximizes structural performance under the given load and boundary conditions. This iteration technique involves solving the equilibrium equations of the finite element method at each step. The design variables $x_e$ are updated using the gradient of the compliance, propagated back to the design variables. 

To prevent checkerboard patterns and mesh dependencies, a filtering method is used to average nearby density values to create a more physical representation. We use density filtering to smooth out any unrealistic patterns and distribute the effect of sensitivity calculations across multiple elements. Each density $x_e$ is updated by the nearby densities $x_i$ in its neighborhood $N_e$. The relative effect that each density value has on the initial density is determined by the weight value $w_{ei}$, Eqn. \eqref{weight fac}. The design density is then updated to the physical density $\tilde{x}_e$ by Eqn. \eqref{density update}. This process, in effect, blurs neighboring density values with each other while maintaining the existing volume fraction. This is designed to discourage small features from forming during the optimization process.

\begin{eqnarray}
w_{ei}=r_{min}-|x_i-x_e|\label{weight fac}\\
\tilde{x}_e = \frac{\sum_{i\in N_e}w_{ei}x_i}{\sum_{i\in N_e}w_{ei}}\label{density update}
\end{eqnarray}

\section{Diffusion Models}\label{sec:3}
Denoising Diffusion Probabilistic Models (DDPMs) are generative models that leverage a probabilistic diffusion process to generate data by iteratively refining samples through a diffusion process. The model is based around a forward diffusion process that gradually adds noise to data, followed by a reverse de-noising process that reconstructs the original data from noisy representations. The forward process $q(x_t|x_{t-1})$ is a probabilistic process that determines the next representation $x_t$ given the previous $x_{t-1}$, by adding a quantity of Gaussian noise to the data, removing information. After the entire forward process over $T$ steps, the representation $x_T$ becomes entirely weighted by the Gaussian noise, resulting in a data representation that does not contain any information about the input data sample $x_0$. This process is defined by a noise schedule $\beta_{1:T}$ which controls the quantity of noise present in the system at a given timestep. The posterior, as presented in \cite{ho2020denoisingdiffusionprobabilisticmodels} then becomes:

\begin{equation}
q(x_t|x_{t-1}) = \mathcal{N}(x_{t-1};\sqrt{1-\beta_t}x_{t-1}, \beta_tI)
\label{diffusionPosterior}
\end{equation}

\begin{equation}
q(x_{1:T}|x_0) = \prod_{t=1}^Tq(x_t|x_{t-1})
\end{equation}

Replicating the reverse process becomes the objective of the DDPM, as a perfect reverse process would reconstruct the original sample without noise. The reverse process is predicted by the network architecture as \cref{diffusionPrior}, which is subject to the model parameters $\theta$. The mean $\mu_\theta$ and standard deviation $\Sigma_\theta$ of the noise are predicted by the model and removed from the sample $x_t$ to produce the previous step $x_{t-1}$.

\begin{equation}
p_\theta(x_{t-1}|x_t) = \mathcal{N}(x_{t-1};\mu_\theta(x_t,t),\Sigma_\theta(x_t,t))
\label{diffusionPrior}
\end{equation}

Diffusion models can be further modified through a conditioning process, where additional contextual or secondary information is integrated into the de-noising process to guide the generation of solutions. Without a conditioning process, the only input into the diffusion model becomes a full sample of noise, which is not selective on the noisy domain. Conditioning involves incorporating a secondary stream of data—often from a different modality, representing constraints, objectives, or external influences—into the latent diffusion model during its iterative denoising steps. This additional information is included with the sample during each step, which allows the information to influence the denoising process. Within the context of design problems, this secondary data stream might include parameters such as material properties, manufacturing limitations, or loading conditions. These inputs serve as constraints or design objectives, allowing the model to generate solutions that not only adhere to performance requirements but also respect practical limitations like manufacturability, cost-efficiency, and compliance with physical laws.

This conditioning process operates by embedding the secondary information alongside the sample, effectively steering the diffusion process toward solutions that meet specified criteria. For instance, the model might incorporate stress distribution maps, volume fraction targets, or boundary conditions as auxiliary data during training. \cite{maze_diffusion_2022, giannone_diffusing_2023} By integrating such context directly into the generative process, the model learns to navigate complex design spaces, balancing competing objectives while adhering to the constraints imposed by the problem. This capability is particularly valuable in topology optimization, where achieving optimal designs requires simultaneous consideration of multiple conflicting factors such as minimizing weight while maximizing stiffness or ensuring manufacturability within additive or subtractive manufacturing processes. The ability to condition on diverse inputs enables latent diffusion models to produce solutions that are practically viable and structurally optimal, representing a significant advancement in computational design methods.

\section{Latent Diffusion Models}\label{sec:4}

\begin{figure*}
    \centering
    \includegraphics[trim={0cm 0cm 0cm 0cm}, clip, width=\linewidth]{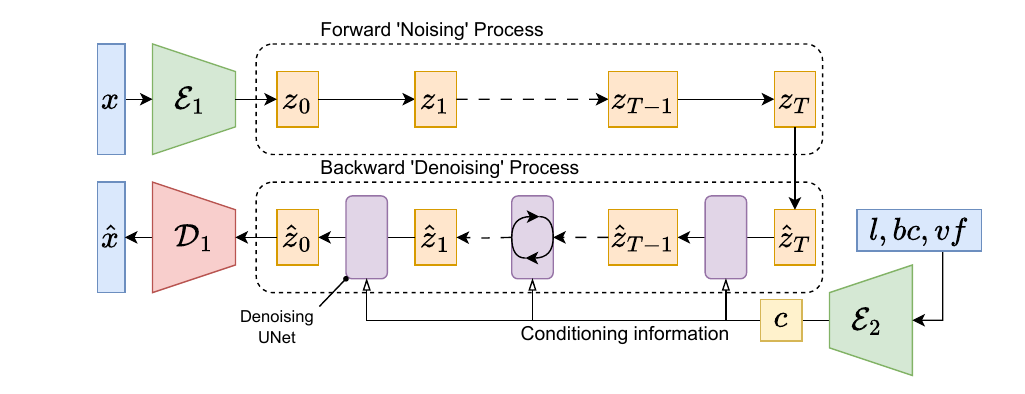}
    \caption{LDM layout. Data passes through the encoder $\mathcal{E}$ (VAE or AE), providing the latent space. Within the latent space, the vector $z$ may be either corrupted with the forward process or reconstructed with the backward process. Conditioning information $l, bc, vf$ are encoded and appended in the latent space for use by the denoising UNet. The decoder $\mathcal{D}$ can be used to remap to the data domain.}
    \label{fig:LDM}
\end{figure*}

Latent Diffusion Models (LDMs) are an extension of DDPMs that operate by gradually transforming random noise into new samples within a low-dimensional latent space. Unlike traditional models that directly operate on high-dimensional data such as images or audio, LDMs use a more computationally efficient approach by modeling the data in a latent space, where the generative process is less costly and more tractable. At the core of this approach is the autoencoder (AE), a neural network architecture that compresses input data into a lower-dimensional latent space. With a sufficiently accurate model, the latent space representation can be considered to be perceptually equivalent to the original domain. The reduced dimensionality of the latent space provides a more computationally efficient space diffusion.

This compression process also enables perceptual compression in which high-frequency details like surface textures or small, non-critical features are discarded to prioritize the representation of globally significant structures and patterns. In the context of topology optimization, features below the manufacturing resolution threshold may be omitted, allowing the model to focus on larger-scale structural characteristics that drive performance. However, this compression introduces a design trade-off: increasing compression reduces the dimensionality of the latent space and improves computational efficiency, but it can also lead to a loss of fidelity in the reconstructed data. The autoencoder's design, such as the size of the latent space and the AE network's architecture, must carefully balance the competing objectives of achieving a compact representation while preserving the critical details necessary for the task at hand. This balance is especially important in LDM-based systems, where the quality of the downstream diffusion model depends on the accuracy of the initial compressed representation. 

In the forward diffusion process of the latent model, an initial latent code $z_0$ corresponding to a data point is progressively corrupted over a series of $T$ time-steps to produce a noisy latent variable $z_t$, where $t \in \{1,2,\dots,T\}$. At each time step, Gaussian noise is added to the latent variable according to the following update rule

\begin{equation}
q(z_t|z_{t-1}) = \mathcal{N}(z_{t-1};\sqrt{1-\beta_t}z_{t-1}, \beta_tI)
\end{equation}

This process causes the latent representation to move gradually from a clean latent vector $z_0 $ to a pure noise latent $z_T$, mirroring the process DDPMs use to map from a clean data sample $x_0$ to a noisy sample $x_T$.

Generation of a new sample is very similar to the traditional diffusion process, where the model learns the reverse of the diffusion process. The process can be initialized by the pure noise vector $z_T$, which requires no prior knowledge of the data distribution of the domain, then repeated over $T$ time-steps until no noise remains in $z_0$. The reverse diffusion process is modeled by a neural network that learns the posterior distribution $p(z_{t-1}|z_t)$, which approximates the true posterior.

The training process of DDPMs and LDMs involves taking any timestep and predicting the noise $\epsilon$ that was added to create that sample. This noise can then be compared to the actual noise that was added in that timestep to return the loss. For LDMs, this can be formulated as:

\begin{equation}
\mathcal{L}_{LDM} = E_{t,z_0,\epsilon} [||\epsilon-\epsilon_\theta(x_t,t)||^2]    
\end{equation}

which is the expected value of the distance between the true noise and predicted noise $\epsilon_\theta$. This loss encourages the model to predict the noise accurately, enabling it to reverse the diffusion process accurately as well.

By performing the diffusion process in a lower-dimensional latent space, LDMs reduce the computational burden compared to working in the original data space. This allows for scalable training and inference, especially in high-dimensional settings like images. Depending on the quality of the AE, the latent representation may be a close substitute for the original data, minimizing the loss in effectiveness incurred by using a latent space. In the case of VAEs, the latent space becomes smooth and dense, which further improves the diffusion model's performance by increasing the size and interpretability of the 'real' sample subset.

\subsection{Autoencoder}\label{sec:4a}
The AE is used in latent diffusion models to generate create a mapping of the input data to and from the latent representation. Autoencoders for image processing map input data $x \in \mathbb{R}^{H\times W\times C}$ to a latent representation $z\in \mathbb{R}^D$ via an encoder $\mathcal{E}$. Similarly, a decoder $\mathcal{D}$ is used to approximate the inverse of the operation such that $\tilde{x} =\mathcal{D}(z)$. The training process involves minimizing a reconstruction loss function, which quantifies the discrepancy between the input to the encoder and the reconstructed output from the decoder, training the two sub-models in parallel. This training can be performed without supervision by minimizing the distance between the actual sample $x$ and reconstructed sample $\tilde{x}$. This distance is the reconstruction loss:
\begin{equation} 
\mathcal{L}_{AE}(\theta, x) = ||\tilde{x}-x||_{2} = ||\mathcal{D}(\mathcal{E}(x)) - x||_{2}
\label{AELoss}
\end{equation}
This drives the AE to learn representations that preserve information, removing redundancy from the input data while still maintaining crucial features for reconstruction.

\subsection{Variational Autoencoder}\label{sec:4b}

\begin{figure*}
    \centering
    \includegraphics[width=1\linewidth]{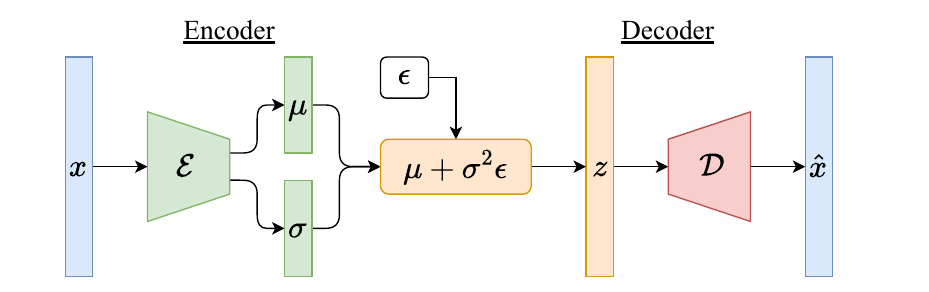}
    \caption{VAE layout. Data passes through the encoder $\mathcal{E}$ to output a mean and standard deviation of the latent space distribution. A resample process maps to a specific latent space vector, which is then decoded by the decoder $\mathcal{D}$}
    \label{fig:VAE}
\end{figure*}

Variational autoencoders (VAEs) are a class of perceptual compression models that model the latent representation as a probabilistic distribution. Rather than performing a one-to-one mapping, VAEs map samples to a distribution of possible values of which the data could be mapped to. For the encoder $\mathcal{E}$, the probabilistic encoder is denoted as $q(z|x)$, which will map to a Gaussian distribution in latent space with some mean $\mu$ and variance $\sigma^2$. The probabilistic decoder $p(x|z)$ maps from some $z$ to a distribution in the sample space.

For a gradient descent algorithm to work, this process must be differentiable. Sampling $z$ from the gaussian distribution, however, is not a differentiable process due to the random component of sampling from a distribution. This step would break the backpropagation of the gradient and prevent the encoder model from receiving any parameter updates. To avoid this, the latent vector is approximated from the gaussian distribution by moving the randomness outside of the process. This is referred to as the \textit{reparameterization trick}, and ensures that the backward process remains differentiable. The latent variable $z$ is expressed as a function of the encoder's output $\mu$ and $\sigma$, and some random noise $\epsilon$ that is external to the model's process. The gradient propagates through the reparameterized variables without needing to pass through the random sampling process $p(\epsilon)$, as the sampling has no learnable parameters. This is expressed as:

\begin{eqnarray}
\tilde{z} = \mu + \sigma^2 \epsilon\\
\epsilon \sim p(\epsilon)
\end{eqnarray}

where $\epsilon$ is drawn from a normal distribution, typically $\mathcal{N}(0, I)$. \cref{fig:VAE} illustrates this relationship as well as the structure of the VAE model's encoder and decoder architecture.

In addition to the reconstruction loss, VAEs also incorporate regularization of the latent space, which encourages the distribution in the latent space to resemble a Gaussian distribution more closely. The KL divergence measures how far the distribution $q(z|x)$ is from the standard normal distribution $\mathcal{N}(0, I)$. The loss function for training a VAE is the sum of the reconstruction loss and KL divergence:

\begin{equation}
    \mathcal{L}_{VAE} = \mathcal{L}_{AE} + \mathcal{L}_{KL}
\end{equation}

where $\mathcal{L}_{KL}$ is the KL divergence between the learned posterior $q(z|x)$ and the assumed prior $p(z)$. This loss term encourages the distributions for each sample to approach the normal distribution, encouraging more overlap in the representations. The higher the overlap, the more structures can be encoded into the same latent space. This regularization prevents the VAE from learning a latent space that maps inputs to sparsely encoded vectors, where the locations of each encoded topology are unrelated to each other. The magnitudes of the latent vectors are compressed around the origin, tending to create a dense distribution that is information rich. Learning an encoded latent distribution rather than an encoded latent vector will give the latent space a dense structure. Each point in the latent space around the assumed prior will correspond to a valid sample from the data distribution. This means that the latent space becomes continuous and smooth, where similar data points in the input space map to nearby points in the latent space. In this structured space, a small change to the latent vector $z$ will lead to small, smooth changes in the generated output $\tilde{x}$. This process regularizes the data because it makes the directions in the latent space more likely to encode meaningful information, rather than a discrete mapping of unrelated topologies.

The KL divergence term, as well as the probabilistic nature of the VAE ensure that the latent space is regularized, informative, and structured. By encouraging the posterior distribution to mimic a Gaussian prior, VAEs promote a latent space that captures the underlying factors that control the variations in the data samples.

\section{\textbf{Methodology}}\label{sec:5}

\subsection{Dataset}\label{sec:5a}
\begin{figure}
    \centering
    \includegraphics[trim={5mm 5mm 5mm 5mm}, clip, width=0.6\linewidth]{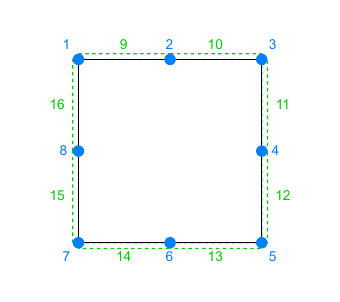}
    \caption{Depiction of the boundary conditions present in the dataset, separated by point conditions (blue) and edge conditions (green)}
    \label{fig:bc sets}
\end{figure}

The dataset used is a generated set of optimized topologies with randomized input conditions. Each sample is generated from a $64 \times 64$ domain of square elements, in which a random element along the boundary is chosen to be the load location. The load is chosen to be a unit magnitude with a direction sampled from $0$ to $2\pi$, sampled from 6 evenly distributed values. Samples are constrained from the nodes along the boundary of the domain. Each sample has between 1 and 4 boundary conditions, chosen from a set of 16 total conditions. 8 of the conditions fix a single point, on the four corners of the domain and the four midpoints of the sides. The remaining 8 conditions are the length of the side that spans between those points. This selection is shown in \cref{fig:bc sets}, demonstrating that the entire edge of the design domain is available for the boundary condition selection process.

Any nodes chosen to be the boundary conditions are fixed in both x and y. Each sample has an associated target volume fraction, ranging from 30\% to 50\% of the full volume of the design domain. The von Mises stress and strain energy density are calculated at the centers of the elements on the initial unoptimized domain as part of the pre-processing of the samples. A unit Young's modulus and Poisson's ratio of 0.3 are used for the calculation of the finite element fields for simplicity, as the magnitudes of these parameters will only affect the scale of the resulting fields, not the relationships. A penalty parameter of 3 and a density filter radius of 2.0 are used. 

The topology optimization algorithm is implemented as described in \cref{sec:02}, using a gradient descent algorithm to perform the minimization of the compliance. A total of 60k samples were generated for the dataset, with rotation and mirroring being used to enhance the effective dataset size.

This setup generates six total channels of information. These are, respectively, the topology channel, the volume fraction, the von-mises stress, the strain energy density, and the load magnitude in the x and y directions. The von-mises stress and strain energy density inherently contain the information of the boundary conditions as a part of the process to create them. The fixed boundary locations will have zero strain, effectively revealing to the model the locations of the boundary conditions. Similarly, the load information is contained in the physical fields as well. This effectively takes the information of the boundary conditions and loading condition, and expands it into a dense plane of information rather than a set of locations. In the context of image generation models, this improves the performance of any subsequent model because the information becomes "dense" rather than the sparsity of only encoding information in the boundary of the domain.

\subsection{Architecture}\label{sec:5b}
We propose two models that are based on the TopoDiff architecture. The first model is a latent diffusional architecture with an autoencoder for latent mapping. The second model uses a variational autoencoder to enhance the smoothness and continuity of the latent space. The autoencoder model, as well as the VAE uses a ConvNext architecture for the encoding of the topology, with a separate pathway for the encoding of the constraints and boundary condition channels. The latent diffusion model is built with a UNet for denoising, and is receives the constraints and boundary conditions as information for guiding the diffusion process.

The VAE is designed to separate the information into two channels to ensure that the information of the conditions (e.g. volume fraction, von-mises stress, strain energy density, and load) is not intertwined with the topology information. To keep the information separate, two distinct autoencoders are trained for the topology $\mathcal{E}_1$ and conditioning information $\mathcal{E}_2$. The LDM is used generatively, so a known topology is not required to initialize the model. It can, however, be initialized with a known set of input conditions. This is what instructs the model to solve a specific design problem.

The topology is encoded into a latent vector $z$ of size $D$, alongside the conditioning vector $c$, also of size $D$. The LDM is then trained with the conditioning vector as a second input to the denoising process, concatenated to the latent vector for the topology. Every step of the diffusion process includes the conditioning information as an input, as shown in \cref{fig:LDM}. This allows the information to be present regardless of the current step. This incorporates the information of the conditions into the UNet without combining the data into the topology channel. During training and sampling of the LDM, noise is only applied to the topology channel and, therefore, is only predicted for the topology channel.

\subsection{Auxiliary Losses}\label{sec:5c}
In the related work Topodiff \cite{maze_diffusion_2022}, a set of conditioning models is used to provide extra information into the denoising process. They utilize two models, a floating material classifier and a compliance regressor. These models are able to directly intervene in the denoising process, ensuring their metrics are minimized alongside the reconstruction loss of the diffusion model. This methodology worked well for the architecture, beating related models in many categories, but requires an expensive training process before the training of the diffusion model. To achieve this process, two additional models need to be trained to predict the compliance and presence of floating material. This training is difficult due to the fact that they need to be used to guide the denoising process, which means they need to be trained on noisy inputs, reducing their effective accuracy. These models predict what the final structure would look like if denoising continued and change the course of the denoising process accordingly. Because these models must predict the compliance of the final structure given a noisy structure, they have inherently lowered accuracy. As a result of this lower accuracy, the guidance process is noisy, especially during the early steps of the diffusion process, where the noise to signal ratio is high.

The inclusion of a VAE in the model architecture gives an opportunity to use more direct conditioning strategies that aren't corrupted by the noise in the diffusion process. Because the VAE is trained to map optimal topologies to a latent space, the set of samples that can be generated from a latent vector is already an 'optimal topology'. The diffusion model only needs to map the topology that best matches the given input conditions. This reduces the need for the compliance regressor, as matching the latent vector via a reconstruction loss is much more likely to produce an optimized design. Similarly, the floating material classifier is also a predictor for a property that is already more present in the latent space than the design space. We can encourage the latent space to be more likely to include these desirable properties via a direct penalization of the VAE.

In addition to using the traditional VAE loss of a reconstruction and KL divergence loss, we propose adding additional loss components for the properties we would like to encourage. For topology optimization, floating material, volume fraction error, load discrepancy, and compliance are all evaluation metrics for a structure's overall efficacy. Volume fraction error, load discrepancy, and floating material presence can all be efficiently calculated, which allows us to penalize the VAE on these properties. We propose the addition of three functions $VF$, $LD$, and $FM$, which calculate these metrics for a topology sample. After the encode decode process, the resulting topology can be passed into each of these evaluation functions to determine the reconstruction's physical realism. We weigh the KL divergence loss with a parameter $\beta_1$ to control how this loss affects the reconstruction accuracy. Similarly, we incorporate these three additional loss functions into the total loss by a weighting factor $\beta_2$ as follows:

\begin{equation} 
\mathcal{L}_{VAE} = \mathcal{L}_{AE} + \beta_1\mathcal{L}_{KL} +\beta_2(VF(\tilde{x}) + LD(\tilde{x}) + FM(\tilde{x}))
\label{VAELoss_with_aux}
\end{equation}

Where:

\begin{eqnarray}
VF(\tilde{x}) = |f - V(\tilde{x})/V_0| \\
LD(\tilde{x}) = 1 - \sum_{i=1}^N\sqrt{(x_i F^x_i)^2 +  (x_iF^y_i)^2}
\end{eqnarray}
The volume fraction loss function $VF(\tilde{x})$ performs a simple calculation of the difference between the sample volume fraction and the actual volume fraction of the sample. The load discrepancy function $LD(\tilde{x})$ calculates the magnitude of the overlap between the element densities $x_i$ and the load vector, $(F^x_i, F^y_i)$.

The floating material function $FM(\tilde{x})$ utilizes the connected components labeling algorithm as implemented in the open-source python module Kornia \cite{eriba2019kornia}. The algorithm assigns a unique integer to each continuous region in a given image. If there are only two unique labels in the image (the topology and the background), then the function will produce a loss of zero. If there are more than two labels, the function will produce a loss of one.

All three auxiliary loss functions are differentiable, allowing them to contribute to the gradients of the model. Given the information in the input conditions vector, the auxiliary losses can be computed along with their derivatives, requiring no additional . We track these metrics over time to see how the performance of the VAE adjusts with the auxiliary losses. In practice, the reconstruction loss will have slightly worse performance in the short term due to the auxiliary losses governing the gradient update slightly. This method, of training the VAE using auxiliary losses, moves the guiding process out of the diffusion model and away from any introduced noise. As a result, no additional models need to be trained and the output of the physical loss functions will always be fully accurate.

\section{\textbf{Results and Analysis}}\label{sec:6}
\begin{figure}
    \centering
    \includegraphics[trim={15mm 5mm 5mm 5mm}, clip, width=0.8\linewidth]{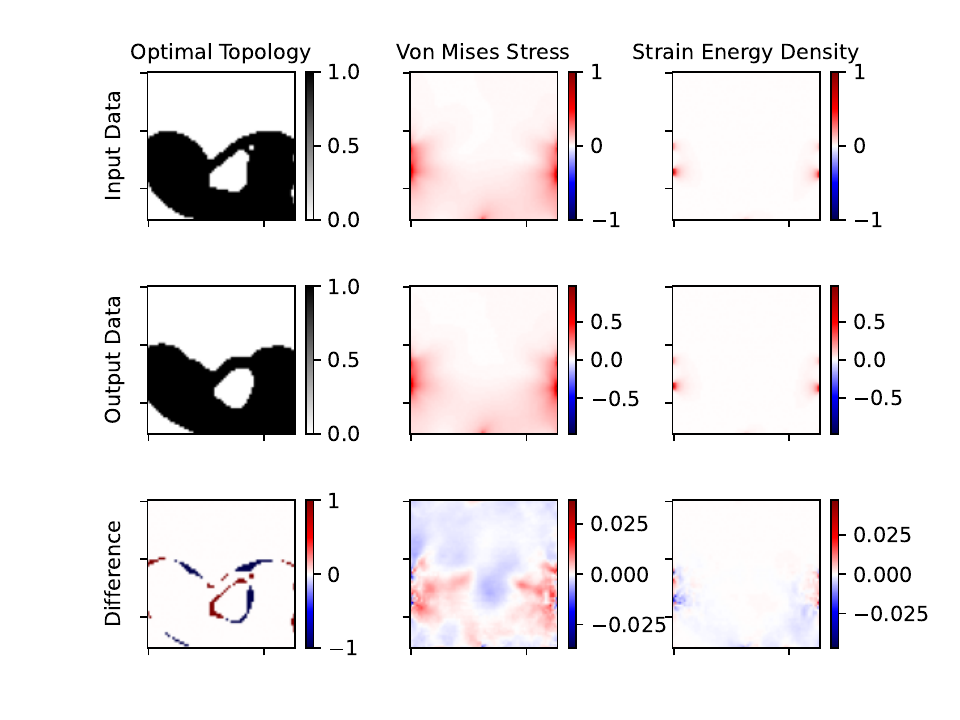}
    \caption{Reconstruction result from the VAE with a latent dimension of 192, $\beta_1$ of 0.075, $\beta_2$ of 0.3. The rows represent the before, after, and error of the VAE reconstruction process. The topology, von Mises stress, and strain energy density channels are .}
    \label{fig:VAE_sample}
\end{figure}

\subsection{Diffusion Model with AE Performance}\label{sec:6a}
We train our AE models for 3,000,000 steps to achieve a convergence in the training loss. The AdamW optimizer is used for the gradient descent method, where the learning rate is set to 1e-4 with a weight decay of 0.05. The batch size is set to 256 to maximize training efficiency for the GPUs used for training, which have a cache of 24 GB.

To assess the influence of control parameters, a hyperparameter study was conducted on the AE. The primary factor that controls the performance of the autoencoder model is the selected compression factor, which is dependent on the latent dimension $D$. For the autoencoder, four latent dimension values of 64, 128, 192, and 256 were chosen to determine the compression factor that contributes to the highest performance downstream diffusion model. To evaluate the performance of the models (AE, VAE, and LDM), 10\% of the dataset is set aside for model evaluation.

\cref{aeLatentDim} contains the reconstruction mean squared error for the various autoencoder models, highlighting the relationship between latent dimension and reconstruction efficiency. Models that are constructed with a larger latent dimension compress the latent space less, resulting in a lower reconstruction loss. While a lower reconstruction loss indicates a latent representation that can be better extracted, it does not account for the smoothness or continuity of the latent space entirely. The diffusion model results may show a different trend in performance than the autoencoder alone.

To review this trend, each autoencoder was used in the training of a downstream diffusion model, which indicates the suitability of the latent space for complex pattern extraction. The results and subsequent compliance analysis of the highest performing of these models is contained in \cref{diffusion_table}, compared with the previous state of the art. 

This autoencoder latent diffusion model architecture has some clear drawbacks with regard to the topology optimization problem. The most glaring issue is the high presence of floating material errors. Because the latent diffusion architecture does not operate on pixel space, it has a harder time maintaining pixel space properties, such as connectivity, volume fraction, and load discrepancy. Additionally, the compliance error outperforms TopologyGAN, but fails to beat the direct diffusion method, TopoDiff. While further training, more hyperparameter optimization, and architecture optimization can improve the results of this architecture, we choose to employ a variational autoencoder with auxiliary losses to target these metrics.


\begin{table}
\centering
\begin{tabular}{
>{\columncolor[HTML]{FFFFFF}}c |
>{\columncolor[HTML]{FFFFFF}}c 
>{\columncolor[HTML]{FFFFFF}}c 
>{\columncolor[HTML]{FFFFFF}}c 
>{\columncolor[HTML]{FFFFFF}}c }
  \textbf{\begin{tabular}[c]{@{}c@{}}Latent\\ Dim $D$\end{tabular}} &
  \textbf{\begin{tabular}[c]{@{}c@{}}F.M.  \\ Loss\end{tabular}}    &
  \textbf{\begin{tabular}[c]{@{}c@{}}L.D.  \\ Loss\end{tabular}}    &
  \textbf{\begin{tabular}[c]{@{}c@{}}V.F.  \\ Loss\end{tabular}}    &
  \textbf{\begin{tabular}[c]{@{}c@{}}Recon.\\ MSE\end{tabular}} \\ \hline
64  & \textbf{1.36\%} & \textbf{0\%} & \textbf{0.51\%} &         1.562E-5  \\
128 &         1.91\%  & \textbf{0\%} &         0.78\%  &         7.342E-6  \\
192 &         2.35\%  & \textbf{0\%} &         0.61\%  & \textbf{3.605E-6} \\
256 &         2.95\%  & \textbf{0\%} &         1.09\%  &         3.871E-6

\end{tabular}
\caption{AE hyperparameter study results for different latent dimensions, displaying floating material loss, load discrepancy loss, volume fraction loss, and reconstruction mean squared error. The best in each category is bold.}
\label{aeLatentDim}
\end{table}

\subsection{Diffusion Model with VAE Performance}\label{sec:6b}
For the variational autoencoder model, a number of other hyperparameters control the model and training process. The two loss weighting parameters $\beta_1$ and $\beta_2$, as well as the size of the latent dimension $D$ are variables which are critical to the overall model performance. To choose the best value of these parameters, we perform a grid search over the possible parameter space. For the variational loss weight $\beta_1$, values of 0.075, 0.15, and 0.3 are used. For the auxiliary loss weight $\beta_2$, values of 0.1 and 0.3 are used. For the latent dimension $D$, values of 64, 128, 192, and 256 are used. These broadly cover the range of usable values for these parameters and provide a set of 24 different runs that may be used for evaluating the model's performance.

 \cref{vaeLatentDim} and \cref{vaeBeta} contain the results for the hyperparameter tuning study. \cref{vaeLatentDim} displays the loss results for the VAE averaged over the latent dimension, while \cref{vaeBeta} displays the loss averaged over the two loss weights. From these results, we can see that the higher latent dimension models produced better losses across all metrics. A latent size of 192 and 256 both produce similar results, with 192 performing better on floating material and reconstruction losses, while 256 has a lower loss for the volume fraction error. For the loss weights, a value of 0.075 $\beta_1$ has the best performance, with a relatively high $\beta_2$ value of 0.3.

A reconstruction sample from the highest performing VAE is shown in \cref{fig:VAE_sample}. The results display the smoothing properties of the VAE, where the high-frequency details are smoothed out of the latent space. The input and output structures are clearly very similar, with minor details missing from the output data. In particular, the small member from the center of the input sample is missing from the output sample, which accounts for a very small change in the structural performance. Similarly, the von Mises stress and strain energy density are accurately encoded and decoded, with small magnitude fluctuations accounting for a majority of the error.


\begin{table}
\centering
\begin{tabular}{
>{\columncolor[HTML]{FFFFFF}}c |
>{\columncolor[HTML]{FFFFFF}}c 
>{\columncolor[HTML]{FFFFFF}}c 
>{\columncolor[HTML]{FFFFFF}}c 
>{\columncolor[HTML]{FFFFFF}}c }
  \textbf{\begin{tabular}[c]{@{}c@{}}Latent\\ Dim $D$\end{tabular}} &
  \textbf{\begin{tabular}[c]{@{}c@{}}F.M.  \\ Loss\end{tabular}}    &
  \textbf{\begin{tabular}[c]{@{}c@{}}L.D.  \\ Loss\end{tabular}}    &
  \textbf{\begin{tabular}[c]{@{}c@{}}V.F.  \\ Loss\end{tabular}}    &
  \textbf{\begin{tabular}[c]{@{}c@{}}Recon.\\ MSE\end{tabular}} \\ \hline
64  &         12.35\%  &         27.80\%  &         1.38\%  &         0.0148  \\
128 &         15.18\%  &         24.02\%  &         1.22\%  &         0.0120  \\
192 & \textbf{11.77\%} & \textbf{17.97\%} &         0.60\%  & \textbf{0.0083} \\
256 &         11.92\%  & \textbf{17.97\%} & \textbf{0.41\%} &         0.0085
\end{tabular}
\caption{VAE hyperparameter study results for different latent dimensions. The best in each category is bold.}
\label{vaeLatentDim}
\end{table}


\begin{table}
\centering
\begin{tabular}{
>{\columncolor[HTML]{FFFFFF}}l 
>{\columncolor[HTML]{FFFFFF}}l |
>{\columncolor[HTML]{FFFFFF}}l 
>{\columncolor[HTML]{FFFFFF}}l 
>{\columncolor[HTML]{FFFFFF}}l 
>{\columncolor[HTML]{FFFFFF}}l }
\textbf{$\beta_1$} &
  \textbf{$\beta_2$} &
  \textbf{\begin{tabular}[c]{@{}l@{}}V.F.  \\ Loss\end{tabular}} &
  \textbf{\begin{tabular}[c]{@{}l@{}}L.D.  \\ Loss\end{tabular}} &
  \textbf{\begin{tabular}[c]{@{}l@{}}F.M.  \\ Loss\end{tabular}} &
  \textbf{\begin{tabular}[c]{@{}l@{}}Recon.\\ MSE\end{tabular}} \\ \hline
0.075 & 0.1 & 0.66\%          & 16.80\%          & 13.30\%          & 0.0079          \\
0.075 & 0.3 & 0.30\%          & \textbf{13.20\%} & \textbf{10.58\%} & \textbf{0.0072} \\
0.15  & 0.1 & 1.27\%          & 23.44\%          & 13.95\%          & 0.0110          \\
0.15  & 0.3 & \textbf{0.21\%} & 23.83\%          & 17.33\%          & 0.0117          \\
0.3   & 0.1 & 2.34\%          & 35.78\%          & 13.70\%          & 0.0185          \\
0.3   & 0.3 & 0.34\%          & 21.68\%          & 11.94\%          & 0.0104         
\end{tabular}
\caption{VAE hyperparameter study results for various loss weightings for $\beta_1$ and $\beta_2$. The best in each category is bold.}
\label{vaeBeta}
\end{table}

While the performance of the diffusion model is dependent on the VAE, a lower reconstruction loss is not the only factor that determines the overall sample quality. The construction of the latent space has a large impact on the diffusion model's overall ability to learn the optimal topology relationships. A model with higher performance on the validation set may indicate a good VAE, but until the downstream diffusion model is trained, the efficacy of the model is still unknown. To achieve the highest downstream accuracy, four models from the hyperparameter tuning study are chosen for further training of the diffusion model. 192 and 256 elements are chosen for the latent size $D$. The variational loss weight $\beta_1$ is chosen to be 0.075, while 0.1 and 0.3 are chosen for the auxiliary loss weight. These parameters provide a set of four models that can be used to achieve high reconstruction accuracy while also further evaluating the auxiliary loss method for topology generation.


\begin{table}
\resizebox{\textwidth}{!}{%
\begin{tabular}{@{}c|cccccc@{}}
\rowcolor[HTML]{C0E6F5} 
\textbf{Models} &
  \textbf{\begin{tabular}[c]{@{}c@{}}Compliance\\ Error (\%)\end{tabular}} &
  \textbf{\begin{tabular}[c]{@{}c@{}}Compliance\\ Error Above \\ 30\% (\%)\end{tabular}} &
  \textbf{\begin{tabular}[c]{@{}c@{}}Median\\ Compliance\\ Error (\%)\end{tabular}} &
  \textbf{\begin{tabular}[c]{@{}c@{}}Floating\\ Material\\ Error (\%)\end{tabular}} &
  \textbf{\begin{tabular}[c]{@{}c@{}}Load\\ Discrepancy\\ (\%)\end{tabular}} &
  \textbf{\begin{tabular}[c]{@{}c@{}}Volume\\ Fraction\\ Error (\%)\end{tabular}} \\ \midrule
\rowcolor[HTML]{FFFFFF} 
TopoGAN       & 48.51         & 10.11         & 2.06          & 46.78         & \textbf{0} & 11.87         \\
\rowcolor[HTML]{FFFFFF} 
TopoDiff UG   & 4.10          & \textbf{2.33} & 0.80          & 6.64          & \textbf{0} & 1.86          \\
\rowcolor[HTML]{FFFFFF} 
TopoDiff G    & 4.39          & 2.56          & 0.83          & \textbf{5.54} & \textbf{0} & \textbf{1.85} \\ \midrule
\rowcolor[HTML]{FFFFFF} 
AE-LDM        & 9.01          & 13.40         & 3.25          & 58.60         & 3.40       & 7.34          \\
\rowcolor[HTML]{FFFFFF} 
VAE-LDM       & 17.39         & 34.44         & 4.15          & 15.00         & 10.00      & 14.84         \\
\rowcolor[HTML]{FFFFFF} 
VAE-LDM w/ AL & \textbf{2.68} & 6.20          & \textbf{0.46} & 7.40          & 3.00       & 2.18         
\end{tabular}%
}\label{diffusion_table}
\end{table}




\begin{figure}[tb!]
    \centering
\subfloat[]{\includegraphics[width = 0.5\linewidth, trim={2.9cm 5cm 2cm 5.2cm}, clip]{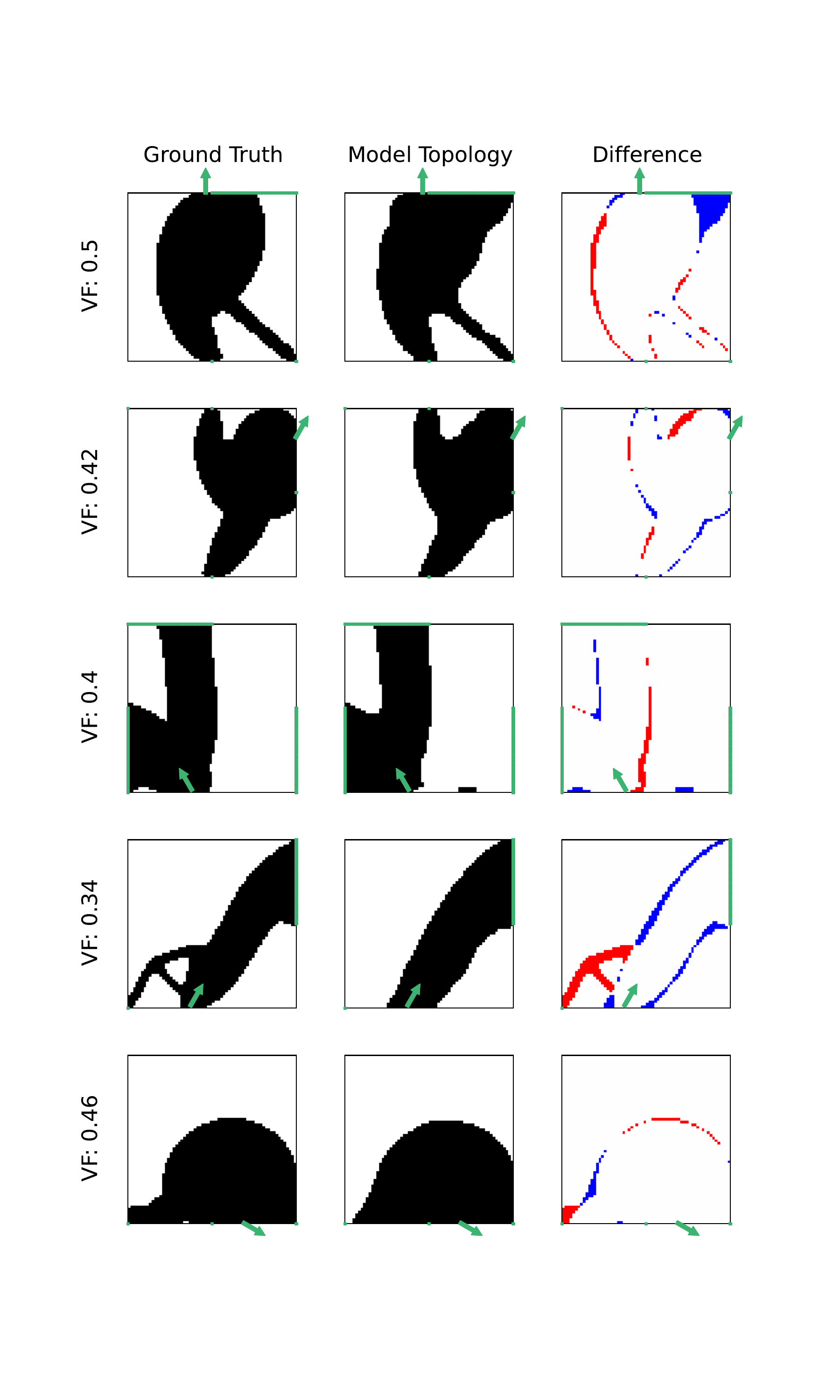}}
\subfloat[]{\includegraphics[width = 0.5\linewidth, trim={2.9cm 5cm 2cm 5.2cm}, clip]{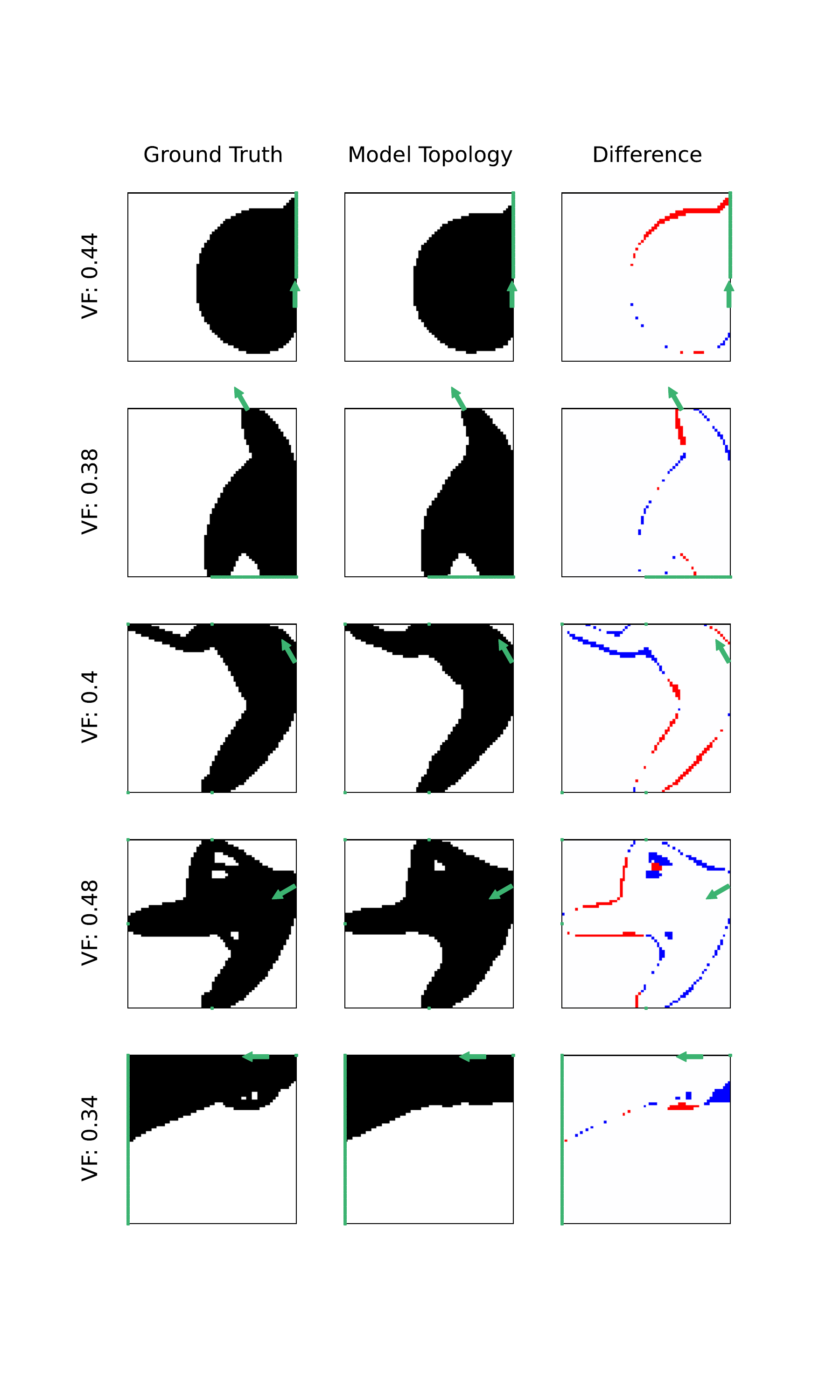}}
\qquad
\caption{Samples from the VAE-LDM, compared to the ground truth topologies. Each row represents a sample and the generated output from the diffusion model, with the target volume fraction labeled. The right column depicts the difference where blue is extra material in the model output and red is missing material from the model output. Boundary conditions are depicted as green line segments or points along the border of the domain. Loads are represented by the arrow in each sample.}
\label{fig:ModelRes}
\end{figure}


\begin{figure}
    \centering
    \includegraphics[width=0.77\linewidth, trim={0.4cm 0cm 1.5cm 1.4cm}, clip]{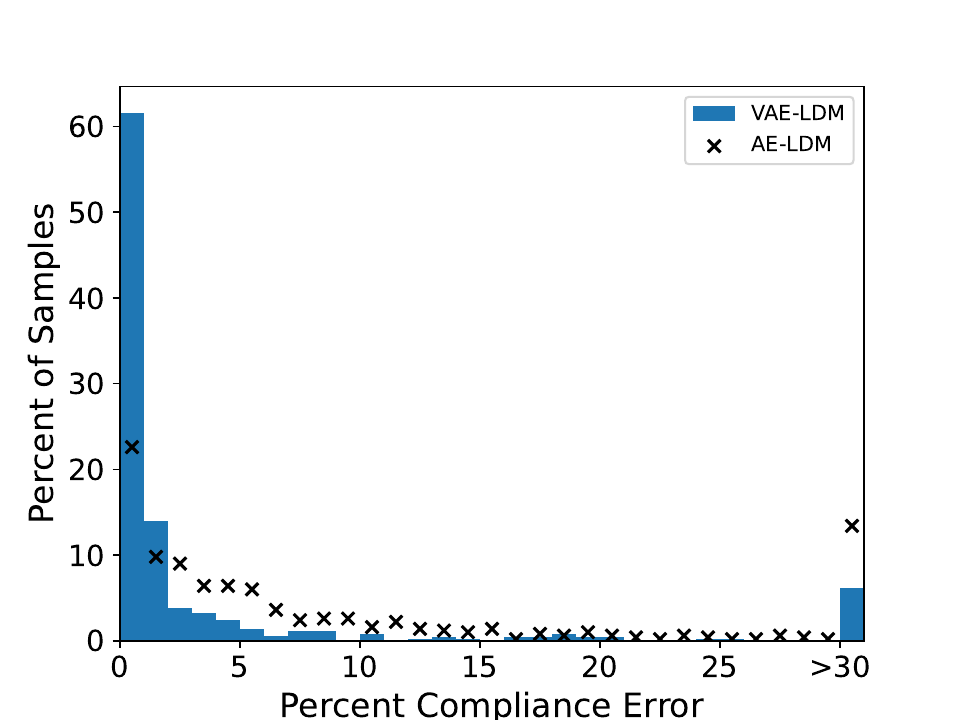}
    \caption{Histogram of samples from the VAE-LDM and AE-LDM based on their relative compliance compared to the ground truth topologies. Samples with more than 30 percent compliance error are pooled into a single data point.}
    \label{fig:ModelHist}
\end{figure}


The framework was tested on a set of validation samples with in-distribution loads and boundary conditions, but out-of-distribution combinations. Results demonstrated that LDM-based optimization achieved designs with improved compliance and material efficiency compared to traditional methods. Computational costs were significantly reduced, as latent space exploration bypassed the high-dimensional domain's iterative sensitivity analyses. Additionally, the model exhibited greater flexibility in incorporating complex constraints, such as symmetry and minimum feature size, directly into the optimization process. 

\cref{fig:ModelRes}a and \cref{fig:ModelRes}b display samples from the full diffusion model compared with their ground truth results. These results, as well as the full analysis reported in \cref{diffusion_table}, clearly demonstrate that VAE-LDMs are capable of matching and surpassing the performance of state-of-the-art diffusion models for topology optimization. Our proposed approach achieves the lowest compliance error percent, given that the high error samples are pruned from the mean calculation. Compared to the reference architecture TopoDiff Guided, the proportion of samples our architecture generates with above 30\% compliance error is about 2.5 times higher (2.56\% to 6.2\%). Despite this, the compliance error of non-defective topologies is in line with expectations, showing a 34\% reduction in compliance error from the highest performing alternative.

Further, the distribution of compliance errors highlights the robustness of our approach. \cref{fig:ModelHist} shows that the distribution of samples is tightly packed around the 1\% compliance error category, indicating a more consistent performance across different load and boundary conditions. This is in contrast to the AE-LDM, which has a wider distribution, with notably more samples in the above 30\% category. The generated topologies exhibit good structural realism, reducing instances of disconnected features, load discrepancies, and volume errors. Our model achieves a 7.4\% floating material rate, outperforming the AE-LDM dramatically, but having slightly lower performance than TopoDiff Guided. The volume fraction error is also significantly reduced compared to the AE-LDM, reducing the 7.34\% error to 2.18\%.

These findings suggest that our approach not only meets but in some cases exceeds the benchmarks set by existing topology optimization methods. While there is room for improvement in reducing the occurrence of high-compliance error cases, the overall efficiency, consistency, and adaptability of the VAE-LDM framework present a compelling case for its application in structural design problems.

\nocite{*}
\bibliographystyle{elsarticle-num}
\bibliography{Diffusion-Model.bib}

\end{document}